\documentclass[reprint,amssymb, amsmath, aps, superscriptaddress, showpacs, prb]{revtex4-1}
\usepackage{amssymb}
\usepackage{amsmath}
\usepackage{xcolor}
\usepackage[colorlinks,bookmarks=false,citecolor=blue,linkcolor=blue,urlcolor=blue]{hyperref}
\usepackage{graphicx}
\usepackage{footmisc,booktabs}
\usepackage{gensymb}
\usepackage{sidecap}
\usepackage{microtype}
\usepackage{epstopdf}

\begin{document}

\title{NMR evidence against a spin-nematic nature of the presaturation phase\\ in the frustrated magnet SrZnVO(PO$_4$)$_2$}

\author{K.~M.~Ranjith}
\email{ranjith.km1857@gmail.com}
\affiliation{Laboratoire National des Champs Magn\'{e}tiques Intenses, LNCMI-CNRS (UPR3228), EMFL, Universit\'{e} \\ Grenoble Alpes, UPS and INSA Toulouse, Bo\^{\i}te Postale 166, 38042 Grenoble Cedex 9, France}
\author{F.~Landolt}
\affiliation{Laboratory for Solid State Physics, ETH Z\"{u}rich, 8093 Z\"{u}rich, Switzerland}
\author{S. Raymond}
\affiliation{Universit\'{e} Grenoble Alpes, CEA, IRIG-MEM-MDN, 38000 Grenoble, France}
\author{A.~Zheludev}
\affiliation{Laboratory for Solid State Physics, ETH Z\"{u}rich, 8093 Z\"{u}rich, Switzerland}
\author{M. Horvati\'{c}}
\email{mladen.horvatic@lncmi.cnrs.fr}
\affiliation{Laboratoire National des Champs Magn\'{e}tiques Intenses, LNCMI-CNRS (UPR3228), EMFL, Universit\'{e} \\ Grenoble Alpes, UPS and INSA Toulouse, Bo\^{\i}te Postale 166, 38042 Grenoble Cedex 9, France}

\date{\today}

\begin{abstract}\noindent
Using $^{31}$P nuclear magnetic resonance (NMR) we investigate the recently discovered presaturation phase in the highly frustrated two-dimensional spin system SrZnVO(PO$_4$)$_2$ [F. Landolt \textit{et al.}, \href{https://dx.doi.org/10.1103/PhysRevB.104.224435}{Phys. Rev. B \textbf{104}, 224435 (2021)}]. Our data provide two pieces of evidence against the presumed spin-nematic character of this phase: \textit{i)} NMR spectra reveal that it hosts a dipolar spin order and \textit{ii)} the $T_1^{-1}$ relaxation rate data recorded above the saturation field can be fitted by the sum of a single-magnon term, exponential in the gap, and a critical second-order term, exponential in the triple gap, leaving no space for a nematic spin dynamics, characterized by a double-gap exponential. We explain the unexpectedly broad validity of the simple fit and the related critical spin dynamics.
\end{abstract}

\maketitle

\section{Introduction}
The possibility of the existence of a purely spin-nematic phase was recognized in 1969 \cite{Blume1969} and discussed theoretically in 1984 \cite{Andreev1984}, but its experimental realizations have been debated only over the last decade, without providing a definite positive conclusion. The point is that such a phase is elusive: it is characterized by a quadrupolar spin order that defies experimental detection \cite{Smerald2015, Smerald2016}, unlike the conventional dipolar spin order that can be observed directly through neutron diffraction or NMR spectroscopy. In particular for spin-$\frac{1}{2}$ systems, a quadrupolar order parameter can be built only from \emph{two}-spin correlators, reflecting the condensation of the two-magnon bound states. Such spin-nematic states have been predicted \cite{Shannon2006, Zhitomirsky2010, Ueda2013} and are sought for in strongly frustrated ferro-antiferromagnetic (ferro-AF) one- and two-dimensional (1D and 2D) spin systems, at high magnetic field close to their saturated phase. Examples include the quasi-1D  LiCuVO$_4$ \cite{Svistov2011, Orlova2017} and the quasi-2D volborthite compound \cite{Ishikawa2015, Yoshida2017, Kohama2019}, in which the putative spin-nematic phase is a presaturation (PS) phase appearing in a \emph{narrow} field range at \emph{very} high magnetic fields, where difficulties of experimental access obstruct precise characterization needed for better identification. That is, as a direct observation of its order parameter is not possible, the identification of a spin-nematic phase necessarily implies accumulation of complementary information from \emph{several} experimental techniques, which is in practice feasible only at moderately high magnetic fields. In this respect, one convenient spin-nematic candidate is BaCdVO(PO$_4$)$_2$ \cite{Nath2008}; however, this quasi-2D compound seems to have a very small magnitude of the order parameter \cite{Povarov2019,Bhartiya2019,Bhartiya2021}. The search for other compounds whose putative spin-nematic phase appears at easily accessible field values is therefore crucial.

Here we address the layered vanadyl phosphate SrZnVO(PO$_4$)$_2$ \cite{Tsirlin2009, Bossoni2011}, a quasi-2D spin system, in which a PS phase was identified very recently above 13.75~T \cite{Landolt2021}. This compound (see Fig.~1 in Ref.~\onlinecite{Landolt2021} for the crystal structure and the spin Hamiltonian) realizes  the frustrated ferro-AF $S = \frac{1}{2}$ Heisenberg model in an approximate $J_1$--$J_2$ square-lattice geometry, prone to exhibit a spin-nematic PS phase when the frustration is strong enough. As compared to its brother compound Pb$_2$VO(PO$_4$)$_2$, in which the absence of spin-nematic character of a very similar PS phase is explained by the weakness of frustration \cite{Landolt2020}, SrZnVO(PO$_4$)$_2$ presents somewhat stronger frustration and much greater quantum fluctuations \cite{Landolt2021}, making it \textit{a priori} a valid spin-nematic candidate.

In order to reveal the real nature of the PS phase in SrZnVO(PO$_4$)$_2$, we carried out a $^{31}$P NMR investigation that is very similar to the one reported previously in Pb$_2$VO(PO$_4$)$_2$ \cite{Landolt2020}, to arrive to the similar conclusion: NMR spectra indicate that the PS phase presents some dipolar spin order, which should be absent in a spin-nematic phase. Furthermore, taking advantage of the lower, more accessible field values, we performed a detailed study of the low-energy (critical) spin dynamics at and above the saturation field $H_{\rm sat}$~= 14.06~T, measured through the nuclear spin-lattice relaxation rate $T_1^{-1}$. These data are fully explained by two contributions, a critical one and a single-magnon one, which match perfectly to general theoretical predictions for a single-magnon condensation, and thus present \emph{no} signature of hypothetical spin-nematic nature. Finally, we discuss the theoretical justification of the simple phenomenological fit to the $T_1^{-1}$ data above $H_{\rm sat}$, as well as the experimentally observed \emph{critical spin dynamics} in comparison to the theoretical prediction \cite{Orignac2007}, which are the most important general (sample-independent) contributions of this work.

\section{Experiment}
NMR experiments were performed on a high-quality single crystal ($\sim$\,4\,$\times$\,2\,$\times$\,0.5~mm$^3$) placed inside the mixing chamber of a $^3$He-$^4$He dilution refrigerator, with the $c$ axis parallel to the applied magnetic field ($H$). The $^{31}$P NMR was measured between 50~mK and 4 K in the magnetic field between 13 and 15 T, using a custom-built spectrometer. The spectra were taken by standard spin-echo sequence and the frequency sweep method. The $T_1^{-1}$ rate was measured by the saturation-recovery method and the time ($t$) recovery of the nuclear magnetization $M(t)$ after a saturation pulse was fitted by the stretched exponential function, $M(t)/M_0 =1-Ae^{-(t/T_1)^{\beta}}$, where $M_0$ is the equilibrium nuclear magnetization, $A \cong 1$ accounts for the imperfection of the excitation (saturation) pulse, and  $\beta$ is the stretch exponent to account for possible inhomogeneous distribution of $T_1^{-1}$  values \cite{Johnston2006, mitrovic2008}. We find $\beta$ values close to 1 (0.92-0.96), indicating a homogeneous system, with the only exception being the near proximity of the phase transition, where the values drop down to $\beta = 0.7$, indicating some smearing (distribution) of the phase boundary.

\begin{figure}[t!]
\includegraphics[width=\columnwidth]{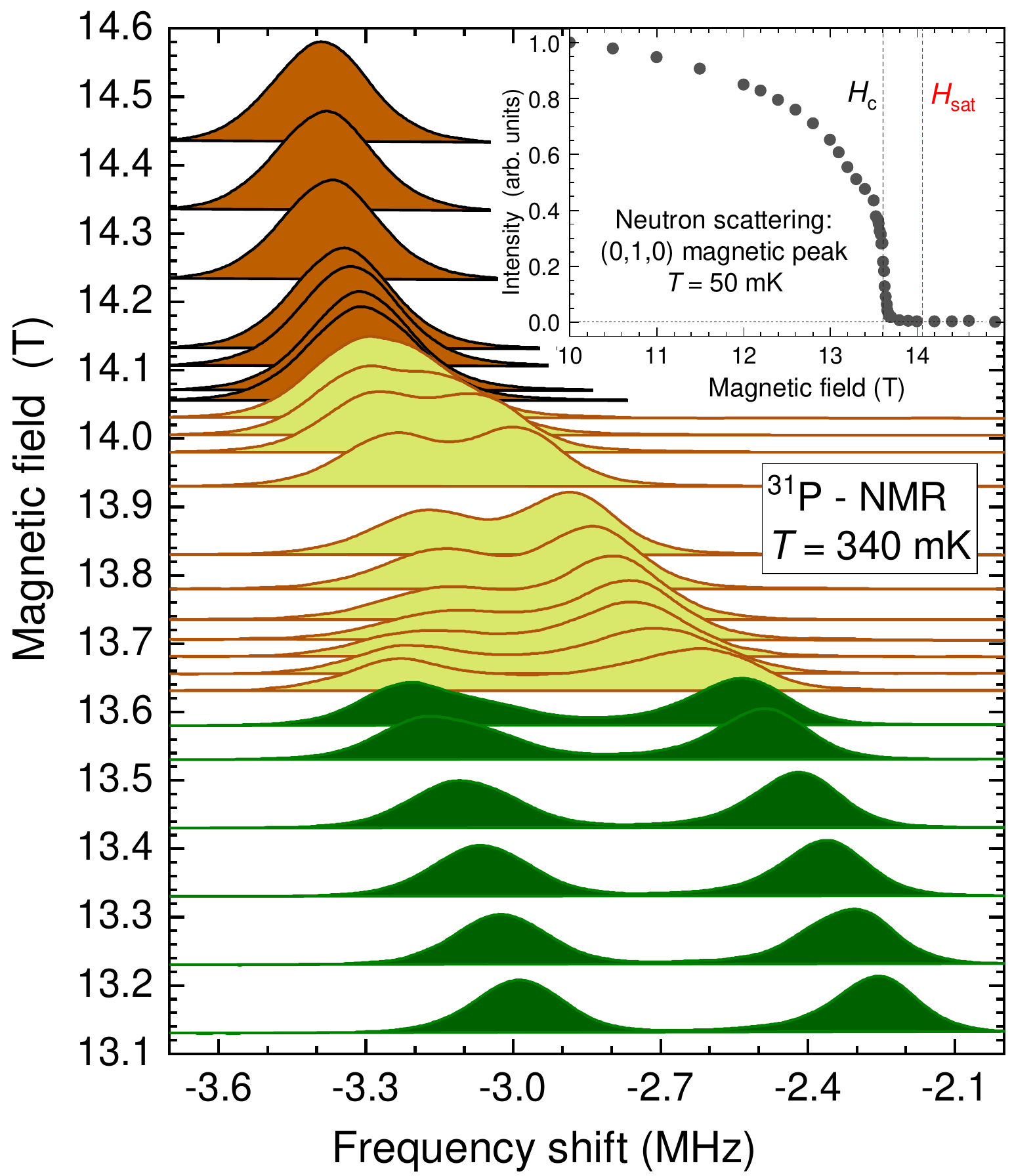}
\caption{$^{31}$P NMR spectra of the P1 site in SrZnVO(PO$_4$)$_2$ measured at $T$~=~340~mK for different magnetic fields applied along the $c$ axis. Different colors indicate the different phases. Inset: Field dependence of the integrated neutron scattering intensity of the $(0,1,0)$ magnetic Bragg reflection.}\label{spectra}
\end{figure}

\section{Results and discussion}
\subsection{NMR spectra}
The observed $^{31}$P NMR spectra are very similar to those observed previously in Pb$_2$VO(PO$_4$)$_2$, presenting two well-separated contributions from the two phosphorous sites present in the system (see Figs. 1 and 5 in Ref.~\onlinecite{Landolt2020}). In this article, we focus on the P1 site that is localized in the planes of V$^{4+}$ spins and is thus a more sensitive probe to the in-plane spin order than the P2 site localized between the planes. Fig.~\ref{spectra} shows the field dependence of the spectrum  measured at 0.34~K and plotted as a function of the frequency shift with respect to the Larmor frequency $^{31}\gamma H$ ($^{31}\gamma$~= 17.236~MHz/T), reflecting the local spin polarization value. Each spectrum is normalized to its integral, offset vertically according to the field value, and color-coded to indicate the three observed phases: \emph{i)} The saturated phase above 14.05~T is homogeneously polarized and is therefore characterized by a single line whose frequency shift tends to a constant saturation value with increasing field. \emph{ii)} In the columnar AF (CAF) phase below 13.6~T we observe a pair of clearly separated lines whose separation measures the staggered moment - the order parameter of this phase, which is also observed by neutron diffraction as the $(0,1,0)$ magnetic Bragg reflection \cite{Landolt2021}. Its field dependence was probed in preliminary neutron diffraction experiments on the IN12 3-axis spectrometer at ILL. As can be seen from inset in Fig.~\ref{spectra}, it survives up to 13.6~T, then disappears in an apparently discontinuous phase transition. \emph{iii)} The intermediate PS phase is characterized by a more complex NMR line-shape, apparently consisting of two overlapping lines, which definitely indicates some dipolar spin order/modulation that is clearly \emph{different} from the one of the CAF phase.

Contrary to what is observed, a spin-nematic phase is expected to be homogeneously polarized, thus presenting the NMR line-shape identical to the one in the saturated phase \cite{Orlova2017}. At the high-field end of the PS phase, the spectrum shrinks continuously into a single line of the saturated phase, indicating a second-order phase transition. At the low-field end of the PS phase, the spectra display a mixed phase in which the two contributions of the PS and the CAF phase overlap, whereas their respective order parameters (line separation) remain essentially unchanged. This is a clear fingerprint of the first-order phase transition, in agreement with neutron data shown in the inset to Fig. \ref{spectra}.

\begin{figure}[t!]
\includegraphics[width=\columnwidth]{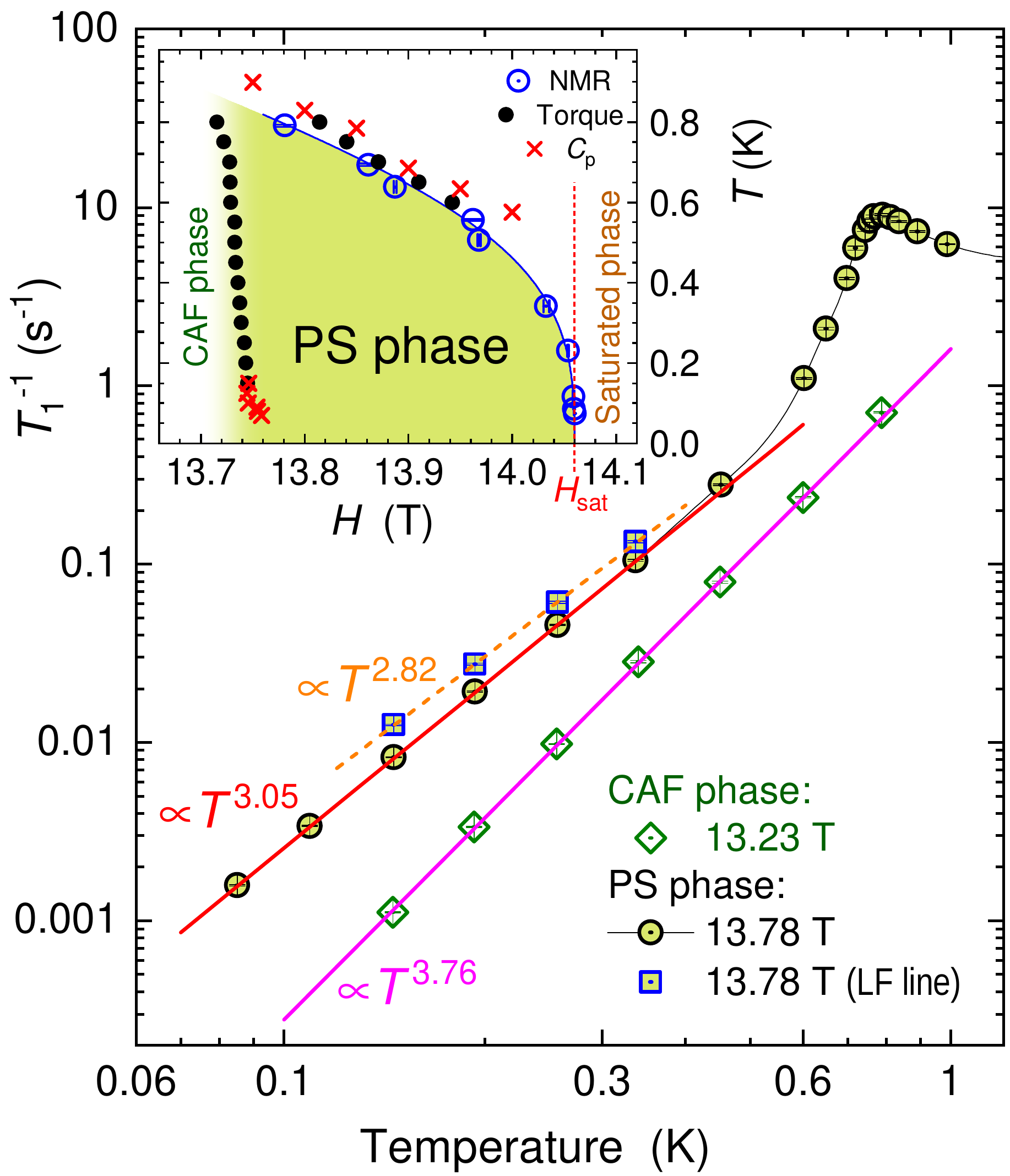}
\caption{Temperature dependence of the NMR $T_1^{-1}$ rate measured at the two field values that characterize the CAF (13.32~T) and the PS (13.78~T) phase. Squares denote data taken at the low-frequency (LF) peak, while other symbols refer to the high-frequency (or unique) peak of the spectra shown in Fig.~\ref{spectra}. Straight lines are power-law fits to low-$T$ data; other lines are guides to the eye. Inset: PS phase diagram determined by NMR from the peak of $T_1^{-1}$ (circles) together with the torque (small dots) and specific-heat (crosses) data from Ref.~\onlinecite{Landolt2021}.}\label{T1Ta}
\end{figure}

\subsection{Spin dynamics}

The spin dynamics, as represented by the $T_1^{-1}$ data shown in Fig.~\ref{T1Ta}, also closely follow the results observed previously in Pb$_2$VO(PO$_4$)$_2$ \cite{Landolt2020}. The second-order phase transition between the ordered PS phase at low temperature and the paramagnetic phase at higher temperature is clearly represented by a peak of $T_1^{-1}$ that reflects the corresponding three-dimensional (3D) critical spin fluctuations. Following the position of this $T_1^{-1}(T)$ peak as a function of magnetic field or, equivalently, the $T_1^{-1}(H)$ peak as a function of temperature defines the NMR data for the phase boundary $T_c(H)$ shown in the inset in Fig.~\ref{T1Ta}, which is in excellent agrement with previously published torque and specific-heat data \cite{Landolt2021}. The low-temperature end of the NMR $T_c(H)$ data provides the estimate of the saturation field value $H_{\rm sat}$~= 14.06~T.

In the ordered phases, the relaxation rate follows a power-law behavior $T_1^{-1}\propto T^{\alpha}$, whose exponent $\alpha$ provides information on their spin dynamics \cite{Beeman1968}. Close to the transition between the two phases, we find $\alpha$(13.23~T)~= 3.8 in the CAF phase and $\alpha$(13.78~T)~= 3.0 in the PS phase. Similar values were found previously in Pb$_2$VO(PO$_4$)$_2$ \cite{Landolt2020} in both phases, $\alpha$(18.7~T)~= 3.4 in the CAF phase and $\alpha$(20.0~T)~= 3.8 in the PS phase, pointing to similar low-energy spin excitations/fluctuations in both phases (close to their interface), despite the difference in their magnetic order.

The $T_1^{-1}$ data in the saturated phase are expected to directly reflect the nature of the relevant spin excitations, corresponding to the nature of the respective PS phase below $H_{\rm sat}$. This means that the single-magnon excitations should be dominant above $H_{\rm sat}$ of a usual one-magnon condensed phase, while the two-magnon (bound-magnon-pair) excitations should be dominant above $H_{\rm sat}$ of a spin-nematic phase. The hallmark of the single-magnon excitations is their gap opening \emph{linearly} with the field, observed through the $T_1^{-1} \propto e^{-\Delta/T}$ dependence, where $\Delta(H) = g\mu_{\rm B}(H-H_{\rm sat})/k_{\rm B}$ is the gap in kelvins, $g$ the Land\'{e} $g$ factor, and $\mu_{\rm B}/k_{\rm B}$~= 0.67171~K/T. The bound magnon pairs are then expected to be recognized by its double-gap, \mbox{$\propto e^{-2\Delta/T}$} contribution, as is reported in volborthite \cite{Yoshida2017, Kohama2019}. However, from the energy diagram relevant to the spin-nematic phase, as shown in Fig.~1 of Ref.~\onlinecite{Zhitomirsky2010}, one can see that the two-magnon gap is smaller than the one-magnon gap only in a \emph{narrow} field range above the (two-magnon) $H_{\rm sat}$ value, whose width is certainly smaller than the width of the nematic phase. Therefore, a possible two-magnon contribution may be observable in the saturated phase only very close to $H_{\rm sat}$, and might be furthermore screened by the corresponding critical fluctuations.

\begin{figure}[t!]
\includegraphics[width=\columnwidth]{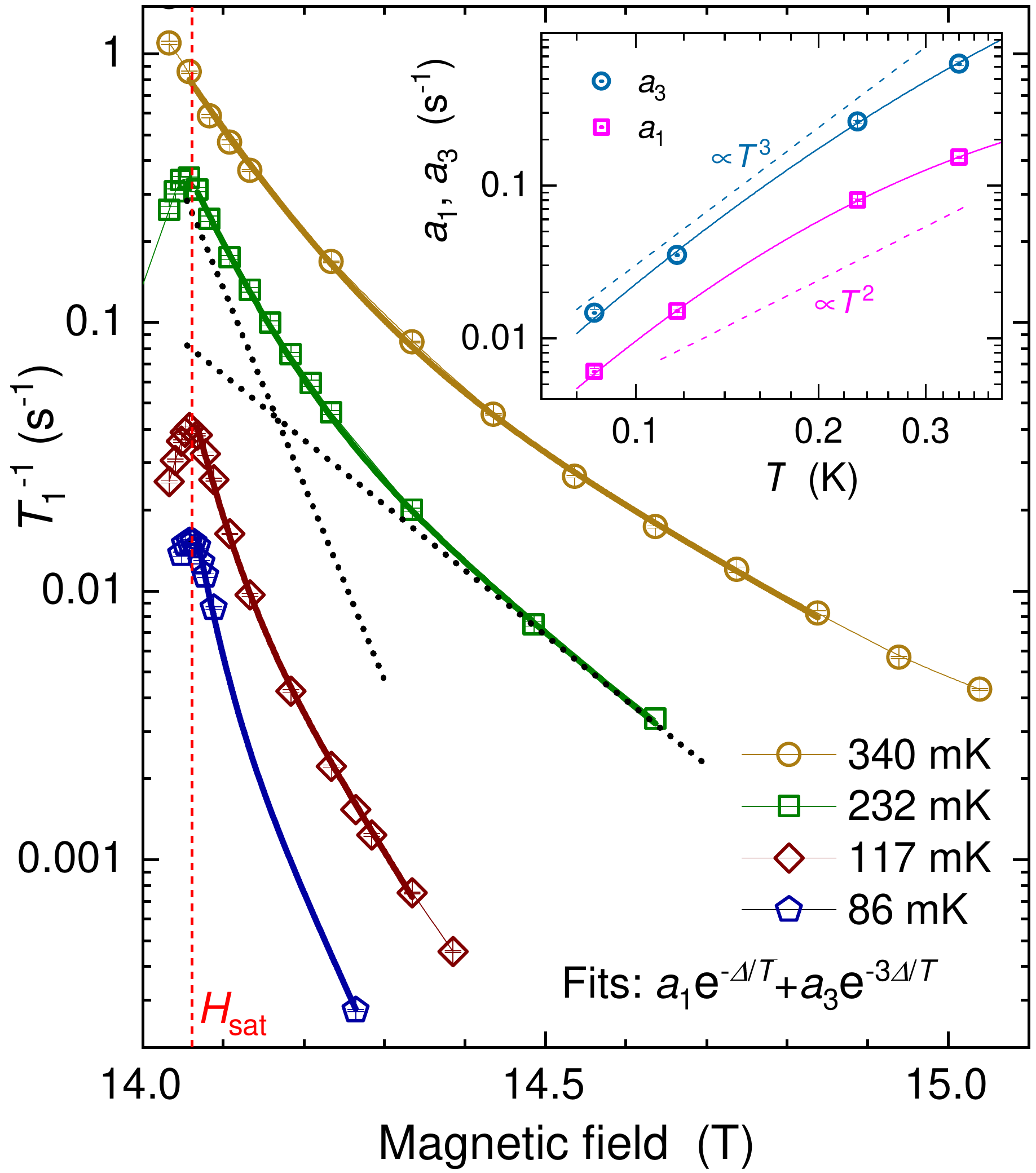}
\caption{Field dependence of the $T_1^{-1}$ rate measured at different temperatures. Thick solid lines show fits using Eq.~(\ref{fit}), while thin lines are guides to the eye. Black dotted lines are the two contributions to the fit at 232~mK. Inset shows the temperature dependence of the two fit parameters, namely the exponential prefactors $a_1$ and $a_3$, the apparent parabolic fits to these data, and the two relevant power-law reference lines.}\label{T1H}
\end{figure}

Figure \ref{T1H} presents the field dependence of the $T_1^{-1}$ data of SrZnVO(PO$_4$)$_2$ covering the field and temperature range relevant for possible detection of two-magnon excitations in the saturated phase. Remarkably, the observed $T_1^{-1}(H)$ dependence above $H_{\rm sat}$ can be fitted by a very simple fit:
\begin{equation}\label{fit}
T_1^{-1}(H,T)=a_1(T) e^{-\Delta(H)/T}+a_3(T) e^{-3\Delta(H)/T},
\end{equation}
where the previously specified gap $\Delta$ is completely defined by the known values of $H_{\rm sat}$, $H$, $T$, and $g$~= 1.926~\cite{Forster2013}, and the \emph{only} two fit parameters are the two amplitudes, $a_1$ and $a_3$, of the single- and triple-gap contributions at the chosen temperature. That is, the two exponentials in Eq.~(\ref{fit}) respectively define the \emph{known}, fixed final and initial slope of the fit, see the black dotted lines in Fig.~\ref{T1H}. These two slopes are directly comparable to the corresponding slopes of the experimental data, and the two fitted amplitudes $a_1$ and $a_3$ cannot change the slopes of the fit but only adjust their ``vertical'' position to make them overlap the experimental data.

The two terms of  Eq.~(\ref{fit}) are respectively associated with the single-magnon excitations and the critical excitations as described by the second order term in Ref.~\onlinecite{Orignac2007}, because both of these excitations are exactly described by the employed exponential terms in their low-$T$ limit, $T \ll \Delta$. In particular, the latter excitations are theoretically described by the self-energy graphs involving \emph{three} magnon propagators (Fig.~2 in Ref.~\onlinecite{Orignac2007}). In order that this three-magnon process can generate a zero-energy spectral density relevant for $T_1^{-1}$ relaxation, the energy conservation imposes that the ingoing magnon must have the energy at least three times bigger than the gap. The corresponding Bose distribution function, when reduced to the Boltzmann factor in the large gap limit ($T \ll \Delta$), then provides the $e^{-3\Delta/T}$ dependence.

Unexpectedly, we find that the fit to the experimental data extends down to $\Delta \rightarrow 0$ ($H \rightarrow H_{\rm sat}$), which is in obvious contradiction with the required low-$T$ limit. In fact, one can \emph{numerically} show that the simple exponential dependence does approximately hold down to $\Delta = 0$, with enough precision to justify the employed fit. This is explained in detail in Appendices \ref{appA} and \ref{appB}, where we first use the numerical solution to establish the relation between the chemical potential $\mu$ and the gap $\Delta$, and then consider the known exact analytical expressions for both contributions to $T_1^{-1}$ as a function of $\mu$.

Having thus established the validity of the employed fit [Eq.~(\ref{fit})] within the description of critical and gapped spin dynamics above the \emph{single}-magnon condensation, we can positively assure the absence of any detectable two-magnon (spin-nematic) contribution. To provide a formal proof for this statement, we add to the fit the third, double-magnon term $a_2 e^{-2\Delta(H)/T}$, and find that the newly fitted curve is nearly identical to the original one: the amplitudes $a_1$ and $a_3$ remain stable within their error bars, while the $a_2$ amplitude is consistent with the zero value. Specifically, the fitted $a_2$ values at 117 and 232~mK are more than one order of magnitude smaller than $a_1$ and, more importantly, they are as much as 11 and 1.7 times smaller than their own respective error bars $\sigma(a_2)$. Even at the highest temperature of 340~mK, where the quality of the fit given by Eq.~(\ref{fit}) starts to slightly degrade (see Fig.~\ref{Fig6} in Appendix~\ref{appB}), this latter $\sigma(a_2)/a_2$ ratio is consequently reduced to 0.84, still clearly compatible with the zero $a_2$ amplitude.

\begin{figure}[t!]
\includegraphics[width=\columnwidth]{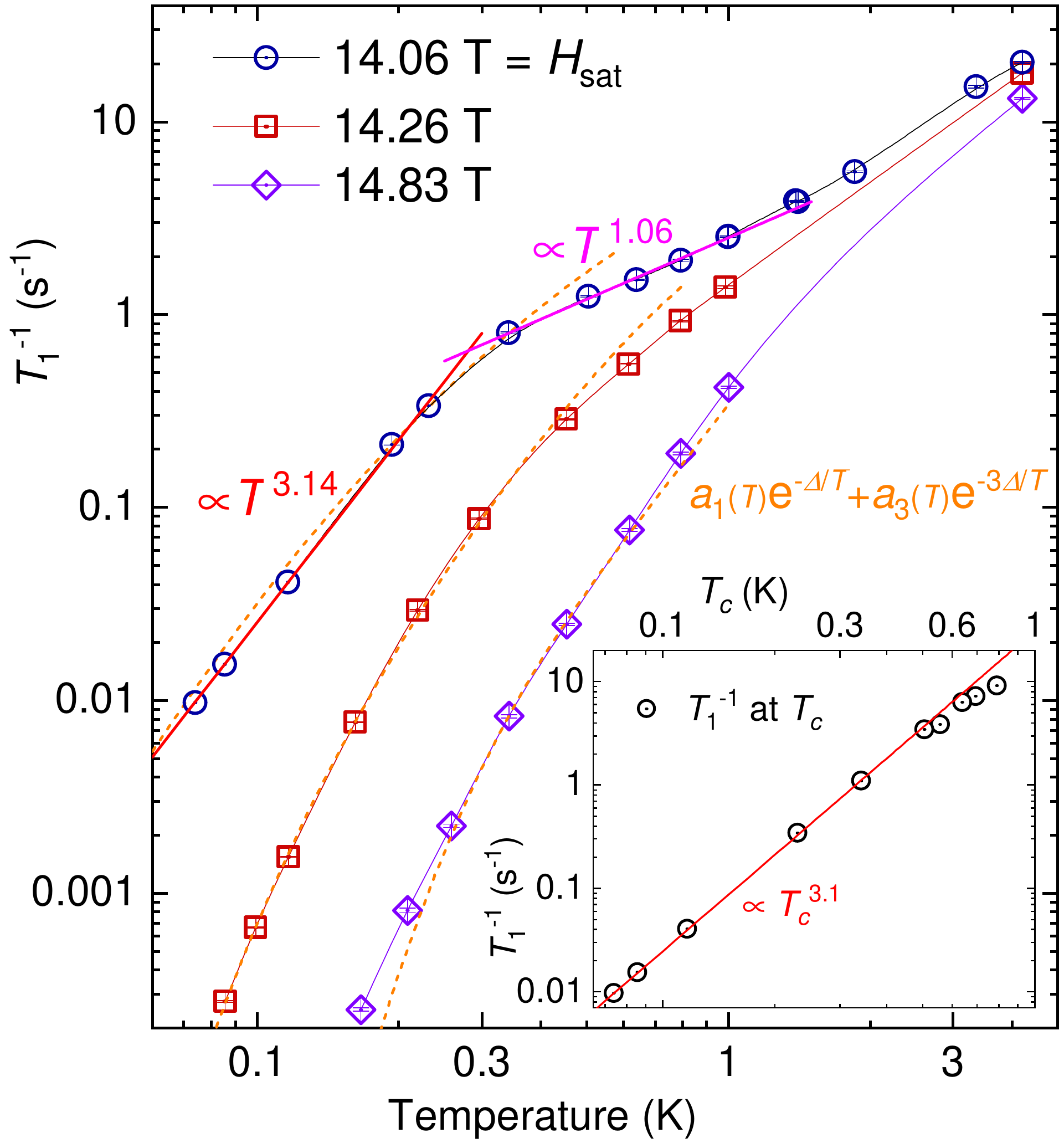}
\caption{Temperature dependence of $T_1^{-1}$ rate measured at several field values at and above the critical field $H_{\rm sat}$~= 14.06~T. Dashed lines are plots using Eq.~(\ref{fit}) and the $a_1(T)$ and $a_3(T)$ dependence shown in the inset to Fig.~\ref{T1H}. Straight lines are power-law fits and thin lines that connect the data points are guides to the eye. Inset shows $T_1^{-1}$ data recorded at the phase boundary, $T_1^{-1}(T_c)$; see also inset to Fig.~\ref{T1Ta}.}\label{T1Tb}
\end{figure}

The temperature dependence of the fitted exponential prefactors $a_1$ and $a_3$ is shown in the inset of Fig.~\ref{T1H}, together with an apparent (on the log-log scale) parabolic fit. Thus obtained $a_1(T)$ and $a_3(T)$ completely define the $T_1^{-1}(H,T)$ dependence of Eq.~(\ref{fit}), which is tested against the measured $T$ dependence of $T_1^{-1}$ data in Fig.~\ref{T1Tb}. The agreement of the simulation and the independent $T_1^{-1}$ data set is remarkable, and furthermore it defines the validity of the proposed fit [Eq.~(\ref{fit})]. Its high-$T$ limit is reached for two reasons: temperature becomes too high as compared to $\Delta$ or/and it reaches value where the (low-$T$) 3D description is no longer applicable, and the system crosses over to the 2D behavior. At $\Delta = 0$, where the formal condition $T \ll \Delta$ is most strongly violated, the simulation remains very good, but is obviously not perfect. Finally, relatively far away above $H_{\rm sat}$, an additional relaxation source is visible at very low temperature.

As regards the critical $T_1^{-1}$ measured exactly at $H_{\rm sat}$, we observe that below 0.2 K its $\propto$\,$T^{3.1}$ dependence is very far from the theoretically expected $\propto$\,$T^{3/4}$ dependence \cite{Orignac2007}. The reason is that here the $H_{\rm sat}$ value is very close to the phase-transition boundary, and the related $T_1^{-1}$ peak is broad enough (see Fig.~\ref{T1H}) so that the peak $T_1^{-1}$ value is practically the same as the $T_1^{-1}(H_{\rm sat})$. In contrast to this, the $T_1^{-1}$ is theoretically expected to diverge to \emph{infinity} at the transition, so that the finite experimental value necessarily results from some kind of the peak broadening, which is \emph{not} described theoretically. Further insight is provided by the inset of Fig.~\ref{T1Tb}, presenting the peak $T_1^{-1}$ data taken exactly at the transition, $T_1^{-1}(T_c)$: we see there that the $T_c^{3.1}$ behavior extends to much higher temperature values, up to $\simeq$0.5~K. Regarding these data, we observe that the boson density at the phase transition $n_c$, in the Hartree-Fock-Popov model, is expected to be linear in $H_{\rm sat} - H$, which was clearly confirmed in the NiCl$_2$-4SC(NH$_2$)$_2$ compound \cite{Paduan-Filho2009, Blinder2017}. This means that $n_c$ is proportional to $T_c^{3/2}$, and the observed peak $T_1^{-1}$ data in SrZnVO(PO$_4$)$_2$ in fact satisfy the $T_1^{-1}(T_c) \propto n_c^2$ relation, a sort of bosonic equivalent of the Korringa law for metallic fermion systems, providing a reasonable explanation for the observed temperature dependence.

Finally, above 0.3~K, as $H_{\rm sat}$ becomes separated enough from the phase boundary line, the $T_1^{-1}(H_{\rm sat},T)$ data crossover to the $T^{1.1}$ dependence, much closer to the theoretically expected $T^{3/4}$; see Appendix \ref{appC}. At higher temperatures, above 1~K, we certainly leave the validity of the 3D approximation, and observe further enhancement of the $T_1^{-1}$ data.

The aforementioned discussion of the low-$T$ dependence $T_1^{-1}(H_{\rm sat},T)$ provides also a natural explanation for the related $a_3(T)$ dependence, shown in the inset of Fig.~\ref{T1H}, being close to $T^3$. As regards the $a_1(T)$ dependence, it is found to be close to the $T^{D-1} = T^2$ dependence that is generally expected for a $D$-dimensional magnon band in the approximation of parabolic dispersion relation (near the edge of the band) \cite{Mukhopadhyay2012}. In fact, a more realistic magnon dispersion results in the \emph{effective} exponent that is somewhat temperature dependent: as shown in the Supplemental Material to Ref.~\onlinecite{Mukhopadhyay2012}, in the 3D regime at low temperature and on decreasing $T$, this exponent is expected to increase across the value of 2, just as is observed in SrZnVO(PO$_4$)$_2$. Altogether, we find that Eq.~(\ref{fit}) provides a complete description of the low-energy spin dynamics in SrZnVO(PO$_4$)$_2$ at and above $H_{\rm sat}$, fully consistent with theoretical expectations.

\section{Summary and conclusion}

In summary, motivated by the possibility that the presaturation phase in SrZnVO(PO$_4$)$_2$ could be a spin-nematic one, and the fact that the corresponding magnetic field values are easily accessible ($H_{\rm sat}$~= 14.06~T), we have performed a detailed NMR study of the related static (NMR spectra) and dynamic ($T_1^{-1}$ relaxation) properties. Both of these provide evidence against the spin-nematic nature of the phase. More importantly, we have provided an extensive set of $T_1^{-1}$ data at and above $H_{\rm sat}$, and fully explained its low-$T$ behavior as a sum of the critical and the one-magnon excitation contribution. The corresponding analysis can be taken as an archetypal description for the one-magnon condensation, which is a necessary reference when searching for NMR signatures of two-magnon condensation. Moreover, we have explained why the experimentally observed low-$T$ limit of the critical $T_1^{-1}(H_{\rm sat},T)$ data must strongly deviate from the formal theoretical prediction $\propto$\,$T^{3/4}$.

\begin{acknowledgments} M.H. acknowledges enlightening exchanges with Martin Klanj\v{s}ek, Edmond Orignac and Mihael Grbi\'{c}. Work at ETH was partially supported by the Swiss National Science Foundation, Division II.
\end{acknowledgments}

\appendix
\section{\label{appA}Chemical potential $\mu$}
The key quantity in describing a BEC-type spin system close to its critical field $H_{\rm sat}$ is the chemical potential $\mu$, which defines the number of excited bosons (magnons) $n_b$, and its relation to the gap $\Delta(H) = g\mu_{\rm B}(H-H_{\rm sat})/k_{\rm B}$ (in kelvins), defined by the applied field. For low density of bosons, their correction to the bare gap is proportional to their number, $\mu =\Delta + 2 U n_b$, thus defining the interaction parameter $U$. Considering that the boson dispersion is parabolic close to the band edge (at $E_0$), which leads to a square-root density of states $g(E) = g_0\,\sqrt{E-E_0}$, the integral that defines $n_b$ can be expressed in terms of the polylogarithm Li$_{3/2}()$ function, leading to the equation \cite{Orignac2007}

\begin{figure}[b!]
\begin{center}
\includegraphics[width=1.0\columnwidth,clip]{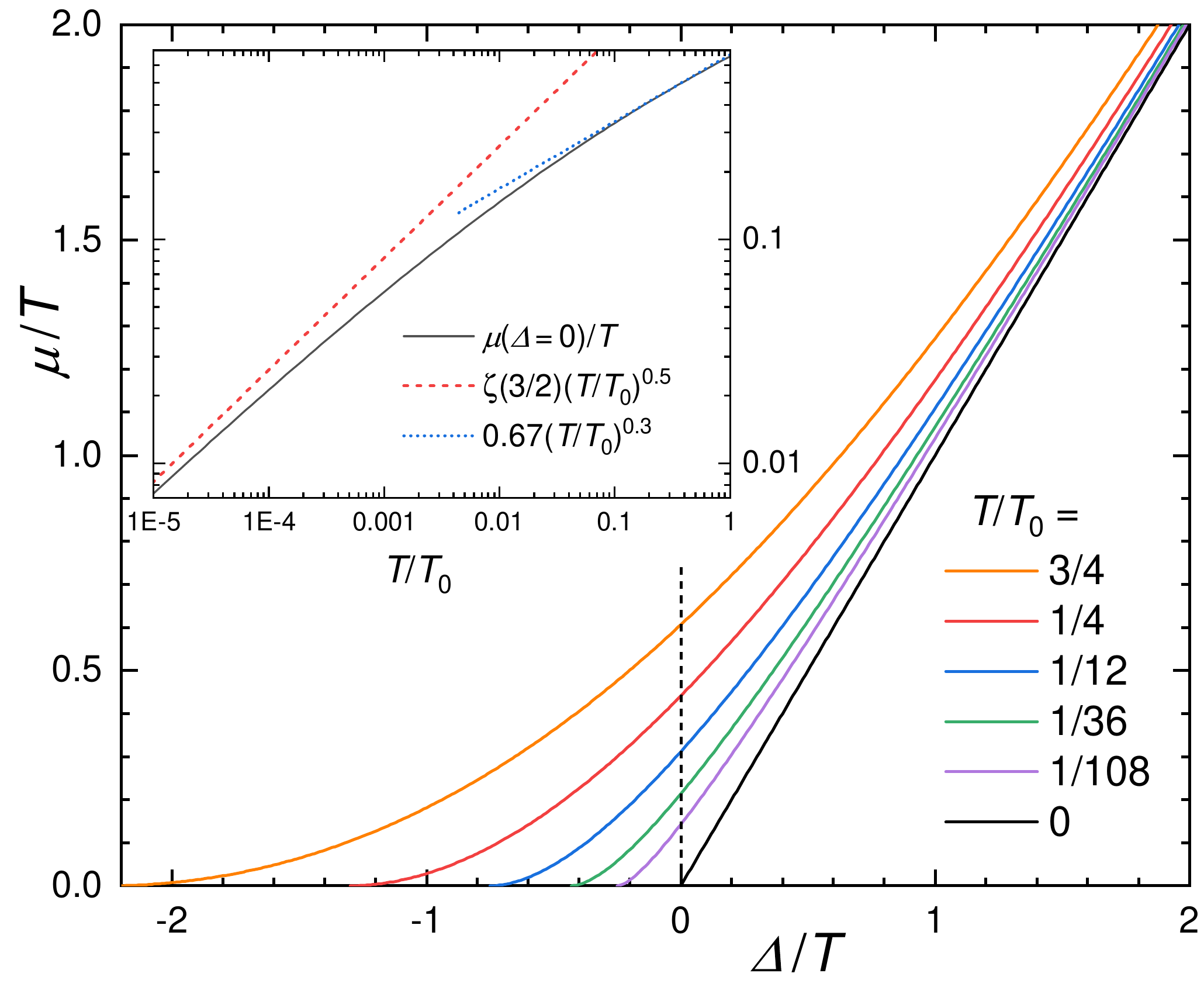}
\caption{The gap (magnetic field) dependence of $\mu$ for an experimentally relevant set of temperatures, obtained by numerical solution of Eq.~(\ref{mu}). Vertical dashed line indicates the critical field (zero gap) position. The temperature dependence of $\mu_c = \mu(\Delta$\,$=$\,$0)$ values taken along this line is shown in the inset, together with its ``relevant'' ($0.05 \lesssim T/T_0 \lesssim 0.5$) power-law approximation $\mu_c \propto T^{1.3}$, as well as the true zero-temperature limit $\mu_c \propto T^{3/2}$.}
\label{Fig5}
\end{center}
\end{figure}

\begin{equation}\label{mu}
    \mu/T = \Delta/T + \sqrt{T/T_0}\,{\rm Li}_{3/2}({\rm e}^{-\mu/T})\,,
\end{equation}
in which the two parameters $U$ and $g_0$ that define the system are contracted into a single one, the characteristic temperature $T_0$.

Equation~(\ref{mu}) is solved numerically to obtain the explicit temperature and field $\mu(T, \Delta)$ dependence, as shown in Fig.~\ref{Fig5}. The inset of this figure focuses on the critical behavior, $\mu_c(T) = \mu(T, \Delta$\,=\,$0)$, where one can see that the convergence toward its mathematical low-$T$ limit $\mu_c(T) \propto T^{3/2}$ is extremely slow and occurs only below $\sim T_0/10^5$, which is totaly unattainable by experiments. Focusing on the experimentally relevant temperature, say $0.05 \lesssim T/T_0 \lesssim 0.5$, we find a significantly different effective exponent, $\mu_c(T) \propto T^{1.3}$.

Finally, should the complete boson/magnon dispersion relations be known, we can numerically calculate the exact density of state that goes beyond its square-root approximation, replace Eq.~(\ref{mu}) by the corresponding numerical equivalent, and thus improve the validity of the computed $\mu(T, \Delta)$ dependence at moderately high temperatures (where $n_b$ is still low enough).

\section{\label{appB}Relaxation rate $T_1^{-1}$}
Dominant contribution to the nuclear spin-lattice relaxation rate $T_1^{-1}$ in the vicinity of the critical field comes from the transverse ($S_+S_-$) spin fluctuations. It has been calculated by Orignac and collaborators \cite{Orignac2007} considering the second-order self-energy graphs. In the relevant zero-frequency limit,
\begin{flalign}
 & \,\,\, 1/T_1^{2nd} \propto T^{3/2}\mathfrak{F}(\mu/T)\, , \,\,\,\,\,\, \textrm{where} \label{T1second} \\
 & \,\,\, \mathfrak{F}(x) = \frac{1}{\sqrt x} \int_{0}^{\infty} \frac{{\rm e}^{-(x+y)}}{[1-{\rm e}^{-(x+y)}]^2} \ln[1+\frac{1-{\rm e}^{-(x+y)}}{{\rm e}^{\frac{x^2}{4y}+x} -1}]\,dy\,. & \nonumber
\end{flalign}
For the large gap $\Delta \gg T$ limit, $\mu \cong \Delta$ and $\mathfrak{F}(x \gg 1) \propto {\rm e}^{-3x}$, leading to the
\begin{equation}\label{3Delta}
   1/T_1^{2nd}(\Delta \gg T) \propto T^{3/2} {\rm e}^{-3\Delta/T}
\end{equation}
dependence, where the triple-gap in the exponent corresponds to a three-magnon scattering process.

The first order process creates the longitudinal ($S_zS_z$) spin fluctuations, which contribute to $T_1^{-1}$ through the off-diagonal components of the hyperfine coupling tensor $A_{zx}$ and $A_{zy}$, naturally generated (at least) by the direct dipolar coupling. For a parabolic magnon dispersion, it is easy to calculate the explicit analytical expression,
\begin{equation}\label{T1first}
    1/T_1^{1st} \propto -T^{2} \ln(1-{\rm e}^{-\mu/T})\,.
\end{equation}
The large gap limit is here obviously
\begin{equation}\label{1Delta}
1/T_1^{1st}(\Delta \gg T) \propto T^{2} {\rm e}^{-\Delta/T}\,,
\end{equation}
corresponding to the Boltzmann limit of the original Bose-Einstein distribution.

\begin{figure}[t!]
\begin{center}
\includegraphics[width=1.0\columnwidth,clip]{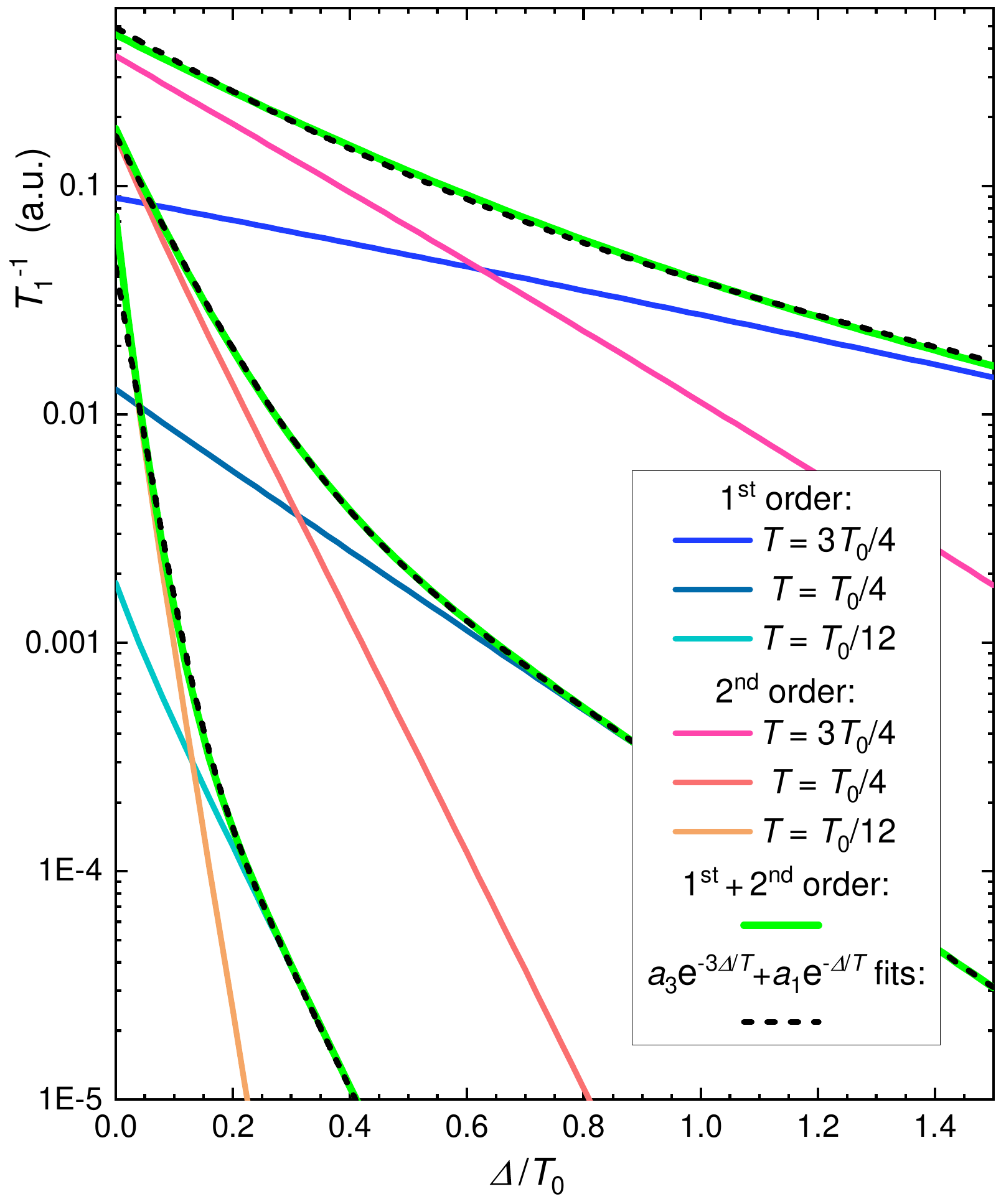}
\caption{Solid lines are theoretically predicted gap/field dependence of the relaxation rates for the first and the second order processes, as defined by Eqs.~(\ref{T1first}), (\ref{T1second}) and (\ref{mu}), for three relevant temperature values. The sum of the two contributions has been fitted by the two-parameter $a_3 e^{-3\Delta/T}+a_1 e^{-\Delta/T}$ fit (dashed lines).}
\label{Fig6}
\end{center}
\end{figure}

While \emph{a priori} both large-gap limits, Eqs.~(\ref{3Delta}) and (\ref{1Delta}), are \emph{not} expected to be valid at small $\Delta/T$, unexpectedly, the $\mu/T$ dependence in Eqs.~(\ref{T1second}) and (\ref{T1first}) coupled to the $\Delta/T$ dependence of $\mu/T$ given by Eq.~(\ref{mu}) extends the \emph{approximate} validity of the large gap limits down to zero gap. This is illustrated in Fig.~\ref{Fig6} for three relevant temperature values, where the relative size of the two contributions to $T_1^{-1}$ (a factor of 5) is chosen so that the figure be qualitatively representative of the measured data presented in Fig.~\ref{T1H}. We see that the theoretical predictions for both contributions are very close to straight lines in the log-lin plots, meaning that they are nearly purely exponential in the gap down to zero. Numerically, we find that this approximation is remarkably precise at $T \simeq T_0/4$, and holds quite well at other experimentally relevant temperatures. Considering that typical error bars on the measured $T_1^{-1}$ data are 2-5\%, the simple single+triple-gap fit given by Eq.~(1) is thus perfectly suitable to describe the magnetic field dependence of experimental relaxation data.

\vspace{-0.5cm}

\section{\label{appC}Critical $T_1^{-1}$}
\begin{figure}[t!]
\begin{center}
\includegraphics[width=1.0\columnwidth,clip]{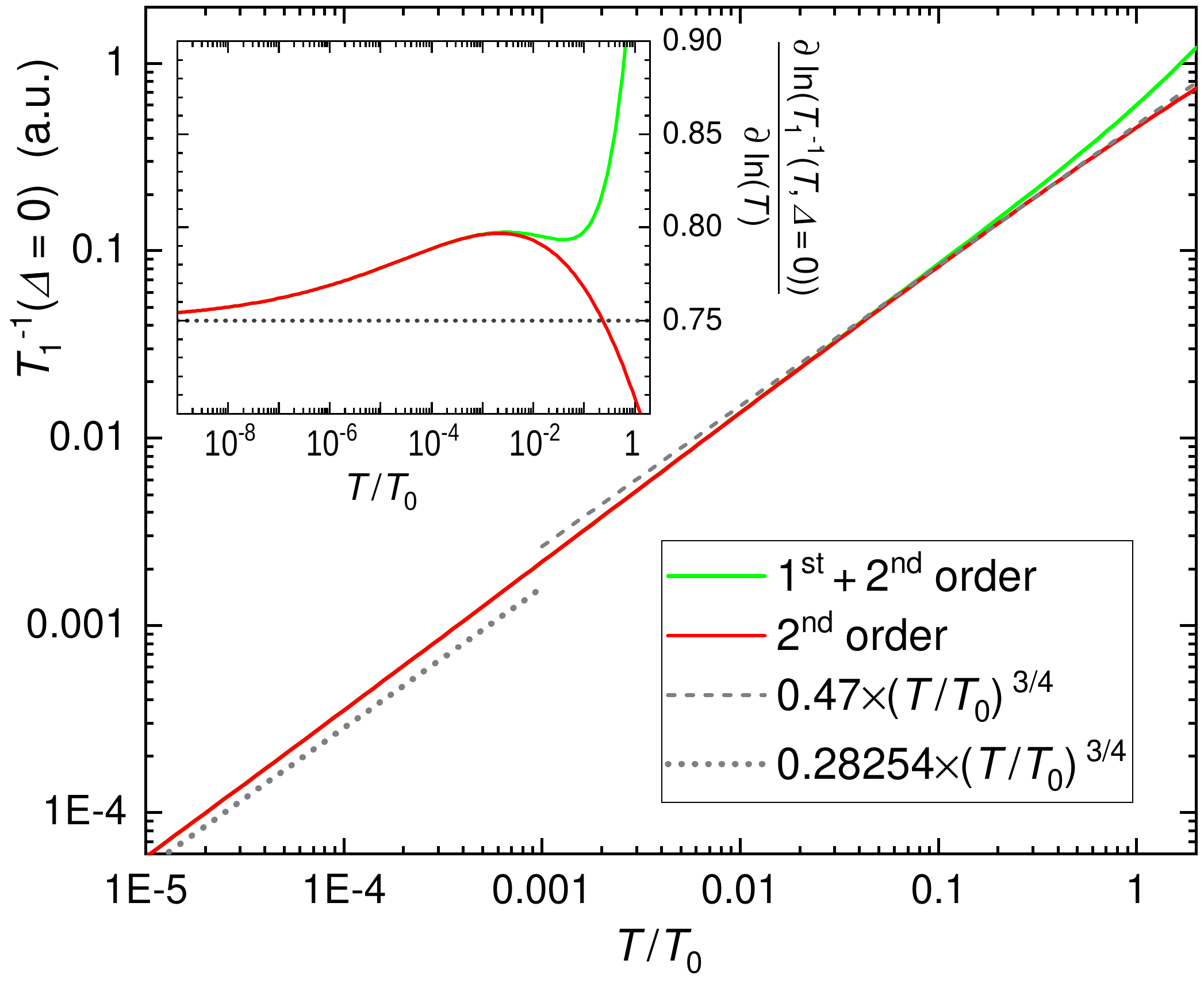}
\caption{Predicted temperature dependence of the critical (i.e., $\Delta$\,=\,0) relaxation rate (solid lines), its low-$T$ limit $T_1^{-1} \propto T^{3/4}$ (dotted line), and the effective power law relevant to the experimentally accessible temperatures (dashed line). Inset shows the temperature dependence of the corresponding effective power-law exponents, $\partial \ln(T_1^{-1}(T, \Delta$\,=\,$0))/\partial \ln(T)$ (solid lines), in the range extended to ultra-low temperatures to show the convergence toward the 3/4 limit (dotted line).}
\label{Fig7}
\end{center}
\end{figure}

In Fig.~\ref{Fig7} we address the theoretically predicted critical behavior, that is, the temperature dependence at zero gap (critical field), in comparison to its $T_1^{-1} \propto T^{3/4}$ limit \cite{Orignac2007}. As expected from very slow convergence of the critical $\mu_c(T)$ at low temperature, the mathematical \mbox{low-$T$} limit of critical $T_1^{-1}$ is likewise reached at extremely low temperatures, and is experimentally totaly inaccessible. Nevertheless, by numerical coincidence, the same 3/4 power exponent is also effectively valid for the second-order contribution $1/T_1^{2nd}$ at an experimentally relevant temperature of $T \simeq T_0/4$. While this cannot be called a universal critical behavior (in the sense of zero-temperature limit), in practice, it does characterize the system. However, it should be noted that in this same temperature range the first-order contribution $1/T_1^{1st}$ is also expected to come into play and thereby increase the effective exponent of the total relaxation rate. Of course, this effect will depend on the relative size of the first- and the second-order contributions; for example, in Fig.~\ref{Fig7} we have considered the same relative size as in Fig.~\ref{Fig6}, leading to experimentally observable exponents of \mbox{0.79-0.90}. Finally, we remark that $T_0$ is also the temperature where the increase of thermally excited boson density is driving the system out of validity of the low-density approximation.

\begin{thebibliography}{99}

\bibitem{Blume1969}
M. Blume and Y. Y. Hsieh, ‘Biquadratic Exchange and
Quadrupolar Ordering, \href{https://doi.org/10.1063/1.1657616}{J. Appl. Phys. \textbf{40}, 1249 (1969)}.

\bibitem{Andreev1984}
A. F. Andreev and I. A. Grishchuk, Spin nematics, \href{http://jetp.ras.ru/cgi-bin/e/index/e/60/2/p267?a=list}{Sov. Phys. JETP \textbf{60}, 267 (1984)}.

\bibitem{Smerald2015}
A. Smerald, H. T. Ueda, and N. Shannon, Theory of Inelastic Neutron Scattering in a Field-Induced Spin-Nematic State,
\href{https://dx.doi.org/10.1103/PhysRevB.91.174402}{Phys. Rev. B \textbf{91}, 174402 (2015)}.

\bibitem{Smerald2016}
A. Smerald and N. Shannon, Theory of NMR $1/T_1$ Relaxation in a Quantum Spin Nematic
in an Applied Magnetic Field, \href{https://dx.doi.org/10.1103/PhysRevB.93.184419}{Phys. Rev. B \textbf{93}, 184419 (2016)}.

\bibitem{Shannon2006}
N. Shannon, T. Momoi, and P. Sindzingre, Nematic Order in Square Lattice Frustrated Ferromagnets, \href{https://dx.doi.org/10.1103/PhysRevLett.96.027213}{Phys. Rev. Lett. \textbf{96}, 027213 (2006)}.

\bibitem{Zhitomirsky2010}
M. E. Zhitomirsky and H. Tsunetsugu, Magnon pairing in quantum spin nematic, \href{https://dx.doi.org/10.1209/0295-5075/92/37001}{Europhys. Lett. \textbf{92}, 37001 (2010)}.

\bibitem{Ueda2013}
H. T. Ueda and T. Momoi, Nematic phase and phase separation near saturation field in frustrated ferromagnets, \href{https://dx.doi.org/10.1103/PhysRevB.87.144417}{Phys. Rev. B \textbf{87}, 144417 (2013)}.

\bibitem{Svistov2011}
L. E. Svistov, T. Fujita, H. Yamaguchi, S. Kimura, K. Omura, A. Prokofiev, A. I. Smirnov, Z. Honda, and M. Hagiwara, New High Magnetic Field Phase of the Frustrated $S = 1/2$ Chain Compound LiCuVO$_4$, \href{https://dx.doi.org/10.1134/S0021364011010073}{JETP Letters \textbf{93}, 21 (2011)}.

\bibitem{Orlova2017}
A. Orlova, E. L. Green, J. M. Law, D. I. Gorbunov, G. Chanda, S. Kr\"{a}mer, M. Horvati\'{c}, R. K. Kremer, J. Wosnitza, and G. L. J. A. Rikken, Nuclear Magnetic Resonance Signature of the Spin-Nematic Phase in LiCuVO$_4$ at High Magnetic Fields,
\href{https://dx.doi.org/10.1103/PhysRevLett.118.247201}{Phys. Rev. Lett. \textbf{118},  247201  (2017)}.

\bibitem{Ishikawa2015}
H. Ishikawa, M. Yoshida, K. Nawa, M. Jeong, S. Kr\"amer, M. Horvati\'{c}, C. Berthier, M. Takigawa, M. Akaki, A. Miyake, M. Tokunaga, K. Kindo, J. Yamaura, Y. Okamoto, and Z. Hiroi, One-Third Magnetization Plateau with a Preceding Novel Phase in Volborthite,
\href{https://dx.doi.org/10.1103/PhysRevLett.114.227202}{Phys. Rev. Lett. \textbf{114}, 227202 (2015)}.

\bibitem{Yoshida2017}
M. Yoshida, K. Nawa, H. Ishikawa, M. Takigawa, M. Jeong, S. Kr\"amer, M. Horvati\'{c}, C. Berthier, K. Matsui, T. Goto, S. Kimura, T. Sasaki, J. Yamaura, H. Yoshida, Y. Okamoto, and Z. Hiroi, Spin dynamics in the high-field phases of volborthite,
\href{https://dx.doi.org/10.1103/PhysRevB.96.180413}{Phys. Rev. B \textbf{96}, 180413(R) (2017)}.

\bibitem{Kohama2019}
Y. Kohama, H. Ishikawa, A. Matsuo, K. Kindo, N. Shannon, and Z. Hiroi, Possible observation of quantum spin-nematic phase in a frustrated magnet,
\href{https://dx.doi.org/10.1073/pnas.1821969116}{Proc. Natl. Acad. Sci. U.S.A. \textbf{116}, 10686 (2019)}.

\bibitem{Nath2008}
R. Nath, A. A. Tsirlin, H. Rosner, and C. Geibel, Magnetic properties of BaCdVO(PO$_4$)$_2$: A strongly frustrated spin-$\frac{1}{2}$ square lattice close to the quantum critical regime, \href{https://dx.doi.org/10.1103/PhysRevB.78.064422}{Phys. Rev. B \textbf{78}, 064422 (2008)}.

\bibitem{Povarov2019}
K.~Yu. Povarov, V.~K. Bhartiya, Z.~Yan, and A.~Zheludev, Thermodynamics of a
	frustrated quantum magnet on a square lattice, \href{https://dx.doi.org/10.1103/PhysRevB.99.024413}{Phys. Rev. B \textbf{99}, 024413 (2019)}.

\bibitem{Bhartiya2019}
V. K. Bhartiya, K. Yu. Povarov, D. Blosser, S. Bettler, Z. Yan, S. Gvasaliya, S. Raymond, E. Ressouche, K. Beauvois, J. Xu, F. Yokaichiya, and A. Zheludev, Presaturation phase with no dipolar order in a quantum ferro-antiferromagnet, \href{https://dx.doi.org/10.1103/PhysRevResearch.1.033078}{Phys. Rev. Research \textbf{1}, 033078 (2019)}.

\bibitem{Bhartiya2021}
V.~K. Bhartiya, S. Hayashida, K.~Yu. Povarov, Z. Yan, Y. Qiu, S. Raymond, and A.
Zheludev, Inelastic neutron scattering determination of the spin  Hamiltonian for BaCdVO(PO$_4$)$_2$, \href{https://dx.doi.org/10.1103/PhysRevB.103.144402}{Phys. Rev. B \textbf{103},  144402  (2021)}.

\bibitem{Tsirlin2009}
A.~A. Tsirlin, B. Schmidt, Y. Skourski, R. Nath, C. Geibel, and H. Rosner, Exploring the spin-$\frac{1}{2}$ frustrated square lattice model with high-field magnetization studies,
\href{https://dx.doi.org/10.1103/PhysRevB.80.132407}{Phys. Rev. B \textbf{80}, 132407 (2009)}.

\bibitem{Bossoni2011}
L. Bossoni, P. Carretta, R. Nath, M. Moscardini, M. Baenitz, and C. Geibel, NMR and $\mu$SR study of spin correlations in SrZnVO(PO$_4$)$_2$: An $S = \frac{1}{2}$ frustrated magnet on a square lattice,
\href{https://dx.doi.org/10.1103/PhysRevB.83.014412}{Phys. Rev. B \textbf{83}, 014412 (2011)}.

\bibitem{Landolt2021}
F. Landolt, Z. Yan, S. Gvasaliya, K. Beauvois, E. Ressouche, J. Xu, and A. Zheludev, Phase diagram and spin waves in the frustrated ferro-antiferromagnet SrZnVO(PO$_4$)$_2$,
\href{https://dx.doi.org/10.1103/PhysRevB.104.224435}{Phys. Rev. B \textbf{104}, 224435 (2021)}.

\bibitem{Landolt2020}
F. Landolt, S. Bettler, Z. Yan, S. Gvasaliya, A. Zheludev, S. Mishra, I. Sheikin, S. Kr\"amer, M. Horvati\'{c}, A. Gazizulina, and O. Prokhnenko, Presaturation phase in the frustrated ferro-antiferromagnet Pb$_2$VO(PO$_4$)$_2$,
\href{https://dx.doi.org/10.1103/PhysRevB.102.094414}{Phys. Rev. B \textbf{102},  094414  (2020)}.

\bibitem{Orignac2007}
E. Orignac, R. Citro, T. Giamarchi, Critical properties and Bose-Einstein condensation in dimer spin systems, \href{https://dx.doi.org/10.1103/PhysRevB.75.140403}{Phys. Rev. B \textbf{75}, 140403(R) (2007)}.

\bibitem{Johnston2006}
D.~C. Johnston, Stretched exponential relaxation arising from a continuous sum of exponential decays, \href{https://dx.doi.org/10.1103/PhysRevB.74.184430}{Phys. Rev. B \textbf{74},  184430  (2006)}.

\bibitem{mitrovic2008}
V. F. Mitrovi\'{c}, M.-H. Julien, C. de Vaulx, M. Horvati\'{c}, C. Berthier, T. Suzuki, K. Yamada, Similar glassy features in the $^{139}$La NMR response of pure and disordered La$_{1.88}$Sr$_{0.12}$CuO$_4$,
\href{https://dx.doi.org/10.1103/PhysRevB.78.014504}{Phys. Rev. B \textbf{78}, 014504 (2008)}.

\bibitem{Beeman1968}
D.~Beeman, P.~Pincus, Nuclear Spin-Lattice Relaxation in Magnetic Insulators,
\href{https://dx.doi.org/10.1103/PhysRev.166.359}{Phys. Rev. \textbf{166}, 359 (1968)}.

\bibitem{Forster2013}
T. F\"orster, F.~A. Garcia, T. Gruner, E.~E. Kaul, B. Schmidt, C. Geibel, and J. Sichelschmidt, Spin fluctuations with two-dimensional XY behavior in a frustrated $S = \frac{1}{2}$ square-lattice ferromagnet,
\href{https://dx.doi.org/10.1103/PhysRevB.87.180401}{Phys. Rev. B \textbf{87},  180401 (2013)}.

\bibitem{Mukhopadhyay2012}
S. Mukhopadhyay, M. Klanj\v{s}ek, M. S. Grbi\'{c}, R. Blinder, H. Mayaffre, C. Berthier, M. Horvati\'{c}, M. A. Continentino, A. Paduan-Filho, B. Chiari, and O. Piovesana, Quantum-Critical Spin Dynamics in Quasi-One-Dimensional Antiferromagnets,
\href{https://dx.doi.org/10.1103/PhysRevLett.109.177206}{Phys. Rev. Lett. \textbf{109}, 177206 (2012)}.

\bibitem{Paduan-Filho2009}
A. Paduan-Filho, K. A. Al-Hassanieh, P. Sengupta, and M. Jaime, Critical Properties at the Field-Induced Bose-Einstein Condensation in NiCl$_2$-4SC(NH$_2$)$_2$,
\href{https://dx.doi.org/10.1103/PhysRevLett.102.077204}{Phys. Rev. Lett. \textbf{102}, 077204 (2009)}.

\bibitem{Blinder2017}
R. Blinder, M. Dupont, S. Mukhopadhyay, M. S. Grbi\'{c}, N. Laflorencie, S. Capponi, H. Mayaffre, C. Berthier, A. Paduan-Filho, and M. Horvati\'{c}, Nuclear magnetic resonance study of the magnetic-field-induced ordered phase in the NiCl$_2$-4SC(NH$_2$)$_2$ compound,
\href{https://dx.doi.org/10.1103/PhysRevB.95.020404}{Phys. Rev. B \textbf{95}, 020404(R) (2017)}.

\end{thebibliography}
\end{document}